\documentclass[12pt]{article}
\usepackage[T1]{fontenc}
\usepackage{graphicx}
\usepackage{amsmath,a4}
\usepackage[hang,footnotesize,bf]{caption2e}

\begin{document}

\begin{center}
{\large \textbf{INTERACTION AND CHAOTIC DYNAMICS OF THE CLASSICAL HYDROGEN
ATOM IN AN ELECTROMAGNETIC FIELD}}\\[0.5cm]
{\large M. Alaburda, V. Gontis and B. Kaulakys}\\[0.5cm]
\emph{Institute of Theoretical Physics and Astronomy, A. Go\v{s}tauto 12,
2600 Vilnius, Lithuania}\\[0.5cm]

\parbox{5in}{\small Expressions for energy and angular momentum changes
of the hydrogen atom due to interaction with the electromagnetic field during
the period of the electron motion in the Coulomb field are derived.
It is shown that only the energy change for the motion between two subsequent passings
of the pericenter is related to the quasiclassical dipole matrix element for transitions between
excited states.}
\end{center}

\section{Introduction}

Classical hydrogen atom in a monochromatic electromagnetic field is one of
the simplest real nonlinear system which dynamics may be regular or chaotic
\cite{dks,kl}, depending on the relative field strength and frequency. Even
one-dimensional classical model of highly excited atom yields results
sufficiently close to the experimental findings. For theoretical analysis
approximate mapping equations of motion, rather than differential equations,
are most convenient \lbrack 2--7\rbrack . Here a two-dimensional map (for
the scaled energy and for relative phase of the field) is generalized for
two-dimensional hydrogen atom, i.e. we calculate energy and angular momentum
changes of the atom interacting with the electromagnetic field.

\section{One-dimensional atom in monochromatic field}

\label{vienmatis}

The Hamiltonian of the hydrogen atom in a linearly polarized monochromatic
electromagnetic field (in atomic units) is \cite{kv,k2}
\begin{equation}
\mathcal{H}=\frac{1}{2}\left( \mathbf{P}+\frac{1}{c}\mathbf{A}\right) ^{2}-%
\frac{1}{r}\;.  \label{h}
\end{equation}
Here $\mathbf{P}$ is the generalized momentum, $c$ is the light velocity,
\begin{equation}
\mathbf{A}=-\frac{c\mathbf{F}}{\omega }\sin (\omega t+\vartheta )\;
\end{equation}
is the vector potential of the field, $\mathbf{F}$, $\omega $ and $\vartheta
$ are the field strength amplitude, field frequency and phase, respectively.
The change of the electron energy can be obtained from the Hamiltonian
equations of motion \cite{ll}
\begin{equation}
\dot{E}=-\mathbf{r}\cdot \mathbf{F}\cos (\omega t+\vartheta )\;.
\label{en_kit}
\end{equation}

One can introduce the scaled energy $E_{s}=E/\omega ^{2/3}$ and the scaled
field strength $F_{s}=F/\omega ^{4/3}$. However, it is convenient [2--7] to
introduce the positive scaled energy $\varepsilon =-2E_{s}$ and the relative
field strength $F_{0}=F_{s}/\varepsilon _{0}^{2}$, with $\varepsilon _{0}$
being the initial scaled energy.

Integration of eq. \eqref{en_kit} over the period of time between two
subsequent passages of the electron at the apocenter results in the change
of the electron energy \cite{gk,kv}
\begin{equation}
\varepsilon _{j+1}=\varepsilon _{j}-\pi F_{0}\varepsilon
_{0}^{2}h(\varepsilon _{j+1})\sin \vartheta _{j}\;,  \label{ev1}
\end{equation}
where
\begin{equation}
h(\varepsilon _{j+1})=\frac{4}{\varepsilon _{j+1}}\mathbf{J}%
_{s_{j+1}}^{\prime }(s_{j+1})\;.  \label{f_h}
\end{equation}
Here $s=\varepsilon ^{-3/2}=\omega /(-2E)^{3/2}$ is the relative frequency
of the field, i.e. the ratio of the field frequency to the electron Kepler
orbital frequency and $\mathbf{J}_{s}^{\prime }(s)$ is the derivative of the
Anger function. Introducing a generating function $G(\varepsilon
_{j+1},\vartheta _{j})$ \cite{ll2,z} one can calculate the phase $\vartheta $
change over the period
\begin{equation}
\vartheta _{j+1}=\vartheta _{j}+2\pi \varepsilon _{j+1}^{-3/2}-\pi
F_{0}\varepsilon _{0}^{2}\eta (\varepsilon _{j+1})\cos \vartheta _{j}\;,
\label{mv1}
\end{equation}
where
\begin{equation}
\eta (\varepsilon _{j+1})=\frac{dh(\varepsilon _{j+1})}{d\varepsilon _{j+1}}%
\;.  \label{eta}
\end{equation}

\begin{figure*}[htb]
\begin{center}
\includegraphics[scale=.55]{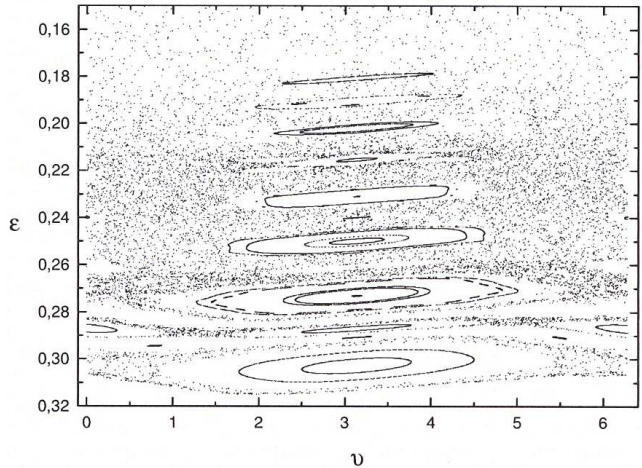}
\end{center}
\par
\vspace{-20pt}
\caption{Trajectories $(\varepsilon,\vartheta)$ for the map
\eqref{ev1} and \eqref{mv1} with the parameter
$\pi F_{0}\varepsilon_{0}^{2}=0.0035$ and initial conditions
$\vartheta_{0}=\pi$, $\varepsilon_{0}=0.3-0.003i$ ($i=0,1,2,\ldots$).}
\label{faz_erdv_did_s}
\end{figure*}

\begin{figure*}[htb]
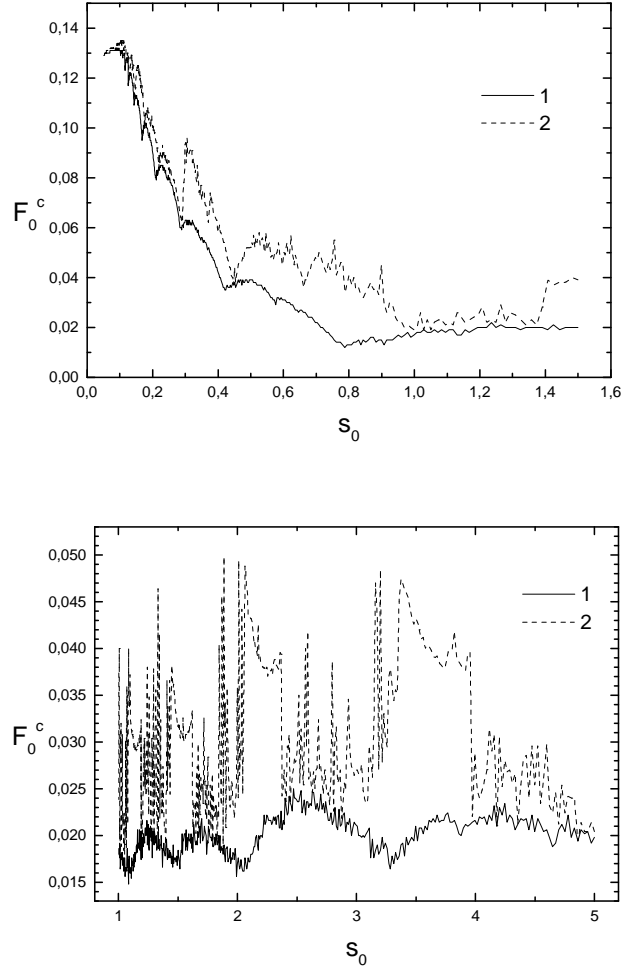

\begin{center}
\includegraphics[scale=.7]{kr_lauk_maziems_s.eps} %
\includegraphics[scale=.7]{kr_lauk_dideliems_s.eps}
\end{center}
\par
\vspace{-20pt}
\caption{Ionization threshold field dependence on the relative frequency
$s_{0}$. Continuous curve represents results calculated with variation of
the initial phase $\vartheta_{0}$, while dotted curve is for
$\vartheta_{0}=0$.}
\label{kr_lauk}
\end{figure*}

Equations \eqref{ev1} and \eqref{mv1} describe the changes of the energy and
phase in time. This map greatly facilitates numerical investigation of
dynamics and ionization process. In figures \ref{faz_erdv_did_s} and figure~%
\ref{kr_lauk} results of the numerical analysis of map \eqref{ev1} - %
\eqref{mv1} are presented.

\section{Two-dimensional atom in monochromatic field}

\label{dvimatis_vienspalvis}

\begin{figure*}[htb]
\begin{center}
\hspace{30pt} \includegraphics[scale=.6]{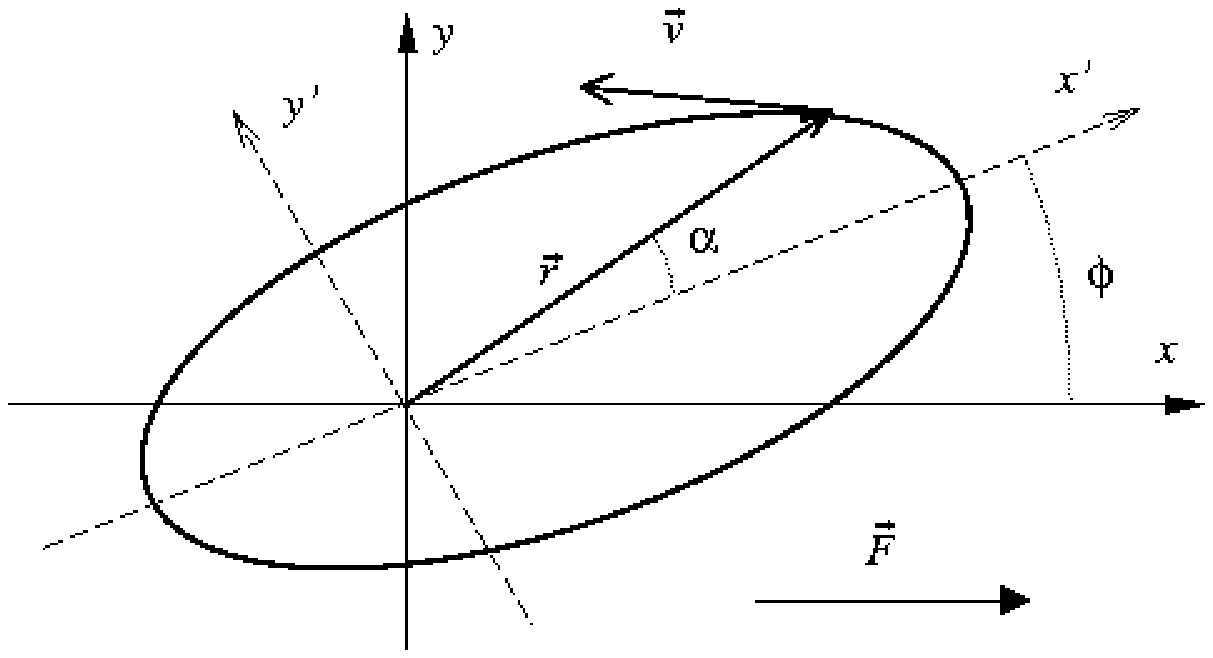}
\end{center}
\par
\vspace{-35pt}
\caption{Two-dimensional atom in the electromagnetic field.}
\label{elipse}
\end{figure*}

For calculation of the energy change of the arbitrary orientated
two-dimensional atom in the electromagnetic field according to equation (3)
one should perform the transformation of the coordinates (figure~\ref{elipse}%
). The change of the angular momentum of the atom it follows from the
Hamiltonian equations of motion
\begin{equation}
\dot{M}=-\frac{\partial \mathcal{H}}{\partial \alpha }=-rF\sin (\alpha
+\varphi )\cos (\omega t+\vartheta )\;.
\end{equation}

By analogy with the scaled energy one can introduce the scaled angular
momentum $\mu =2M_{s}=2M\omega ^{1/3}$. Moreover, it is convenient to
introduce the parametric equations of motion for the Coulomb potential
[6--8].

Integration of equations (3) and (8) for the half of the period of the
electron motion, i.e. for transition from apocenter to pericenter yield the
scaled energy and angular momentum changes in the linearly polarized
electromagnetic field
\begin{multline}
\varepsilon _{j+1}=\varepsilon _{j}+\frac{2\pi F_{0}\varepsilon _{0}^{2}}{%
\varepsilon _{j+1}}\left\{ -\left[ \mathbf{J}_{s}^{\prime }(z)\sin \vartheta
_{j}+\mathbf{E}_{s}^{\prime }(z)\cos \vartheta _{j}\right] \cos \varphi
\right.  \\
\left. +\frac{\sqrt{1-e^{2}}}{e}\left[ \left( \mathbf{J}_{s}(z)-\frac{\sin
(s\pi )}{s\pi }\right) \cos \vartheta _{j}-\left( \mathbf{E}_{s}(z)-\frac{%
1-\cos (s\pi )}{s\pi }\right) \sin \vartheta _{j}\right] \sin \varphi
\right\} \;  \label{e0pi}
\end{multline}
and
\begin{multline}
\mu _{j+1}=\mu _{j}+\frac{2\pi F_{0}\varepsilon _{0}^{2}}{\varepsilon _{j+1}}%
\left\{ \frac{\sqrt{1-e^{2}}}{e}\left[ \left( \mathbf{J}_{s}(z)-\frac{\sin
(s\pi )}{s\pi }\right) \sin \vartheta _{j}\right. \right.  \\
\left. \left. +\left( \mathbf{E}_{s}(z)-\frac{1-\cos (s\pi )}{s\pi }\right)
\cos \vartheta _{j}\right] \cos \varphi +\left[ \left( -\mathbf{J}%
_{s}^{\prime }(z)+(1+e)\frac{\sin (s\pi )}{\pi }\right) \cos \vartheta
_{j}\right. \right.  \\
\left. \left. +\left( \mathbf{E}_{s}^{\prime }(z)+(1+e)\frac{\cos (s\pi )}{%
\pi }+\frac{1-e}{\pi }\right) \sin \vartheta _{j}\right] \sin \varphi
\right\} \;.  \label{m0pi}
\end{multline}

Here $\mathbf{E}_{s}(z)$is the Weber function, $z=es$ and e is the
eccentricity of the ellipsis. By analogy or from equations (9) and (10)
choosing appropriate initial phases of the field one can calculate the
energy and angular momentum changes for the electron motion from pericenter
to apocenter as well as for the complete period with different initial
conditions.

So, for motion between two subsequent passages at the apocenter we have
generalization of eq. (4) for the energy change
\begin{equation}
\varepsilon _{j+1}=\varepsilon _{j}+\frac{4\pi F_{0}\varepsilon _{0}^{2}}{%
\varepsilon _{j+1}}\left\{ -\mathbf{J}_{s}^{\prime }(z)\sin \vartheta
_{j}\cos \varphi +\frac{\sqrt{1-e^{2}}}{e}\left[ \mathbf{J}_{s}(z)-\frac{%
\sin (s\pi )}{s\pi }\right] \cos \vartheta _{j}\sin \varphi \right\} \;
\label{e-pipi}
\end{equation}
and the angular momentum change
\begin{multline}
\mu _{j+1}=\mu _{j}+\frac{4\pi F_{0}\varepsilon _{0}^{2}}{\varepsilon _{j+1}}%
\left\{ \frac{\sqrt{1-e^{2}}}{e}\left[ \mathbf{J}_{s}(z)-\frac{\sin (s\pi )}{%
s\pi }\right] \sin \vartheta _{j}\cos \varphi \right.  \\
+\left. \left[ -\mathbf{J}_{s}^{\prime }(z)+(1+e)\frac{\sin (s\pi )}{\pi }%
\right] \cos \vartheta _{j}\sin \varphi \right\} \;.  \label{m-pipi}
\end{multline}

One can calculate the energy and angular momentum changes for the electron
motion between two subsequent passages at the pericenter in the similar way.

\subsection{Approximation for relatively high frequency $s$}

For relatively high frequency of the field, $s\gg 1$, the asymptotic form of
the Anger function $\mathbf{J}_{s}(se)$ and its derivative $\mathbf{J}%
_{s}^{\prime }(se)$ may be used \cite{k2}. Then eq. (11) may be written in
the form
\begin{multline}
\varepsilon _{j+1}=\varepsilon _{j}+\pi F_{0}\varepsilon _{0}^{2}\left\{
(e-2)\left[ 4b-\frac{4a}{5}\varepsilon _{j+1}-\frac{\sin (\varepsilon
_{j+1}^{-3/2}\pi )}{\pi }\varepsilon _{j+1}^{2}\right] \sin \vartheta
_{j}\cos \varphi \right.  \\
+\frac{\sqrt{1-e^{2}}}{e}\left[ 4a\varepsilon _{j+1}^{-1/2}-\frac{2\sin
(\varepsilon _{j+1}^{-3/2}\pi )}{\pi }\varepsilon _{j+1}^{1/2}-\frac{2b}{35}%
\varepsilon _{j+1}^{3/2}\right.  \\
\left. \left. +(e-1)\left( 4b\varepsilon _{j+1}^{-3/2}-\frac{4a}{5}%
\varepsilon _{j+1}^{-1/2}-\frac{\sin (\varepsilon _{j+1}^{-3/2}\pi )}{\pi }%
\varepsilon _{j+1}^{1/2}\right) \right] \cos \vartheta _{j}\sin \varphi
\right\} \;.  \label{en_skl}
\end{multline}
Introducing the generating function we can obtain the iterative equation for
the phase $\vartheta _{j}$ as a generalization of eq. (6)
\begin{multline}
\vartheta _{j+1}=\vartheta _{j}+2\pi \varepsilon _{j+1}^{-3/2}+\pi
F_{0}\varepsilon _{0}^{2}\left\{ (2-e)\left[ \frac{4a}{5}-\frac{3\cos
(\varepsilon _{j+1}^{-3/2}\pi )}{2}\varepsilon _{j+1}^{-1/2}\right. \right.
\\
\left. +\frac{2\sin (\varepsilon _{j+1}^{-3/2}\pi )}{\pi }\varepsilon
_{j+1}\right] \cos \vartheta _{j}\cos \varphi +\frac{\sqrt{1-e^{2}}}{e}%
\left[ 2a\varepsilon _{j+1}^{-3/2}-3\cos (\varepsilon _{j+1}^{-3/2}\pi
)\varepsilon _{j+1}^{-2}\right.  \\
+\frac{\sin (\varepsilon _{j+1}^{-3/2}\pi )}{\pi }\varepsilon _{j+1}^{-1/2}+%
\frac{3b}{35}\varepsilon _{j+1}^{1/2}+(e-1)\left( 6b\varepsilon
_{j+1}^{-5/2}\right.  \\
\left. \left. \left. -2a\varepsilon _{j+1}^{-3/2}-\frac{3\cos (\varepsilon
_{j+1}^{-3/2}\pi )}{2}\varepsilon _{j+1}^{-2}+\frac{\sin (\varepsilon
_{j+1}^{-3/2}\pi )}{2\pi }\varepsilon _{j+1}^{-1/2}\right) \right] \sin
\vartheta _{j}\sin \varphi \right\} \;.  \label{faz_skl}
\end{multline}

\begin{figure*}[htb]
\begin{center}
\includegraphics[scale=.55]{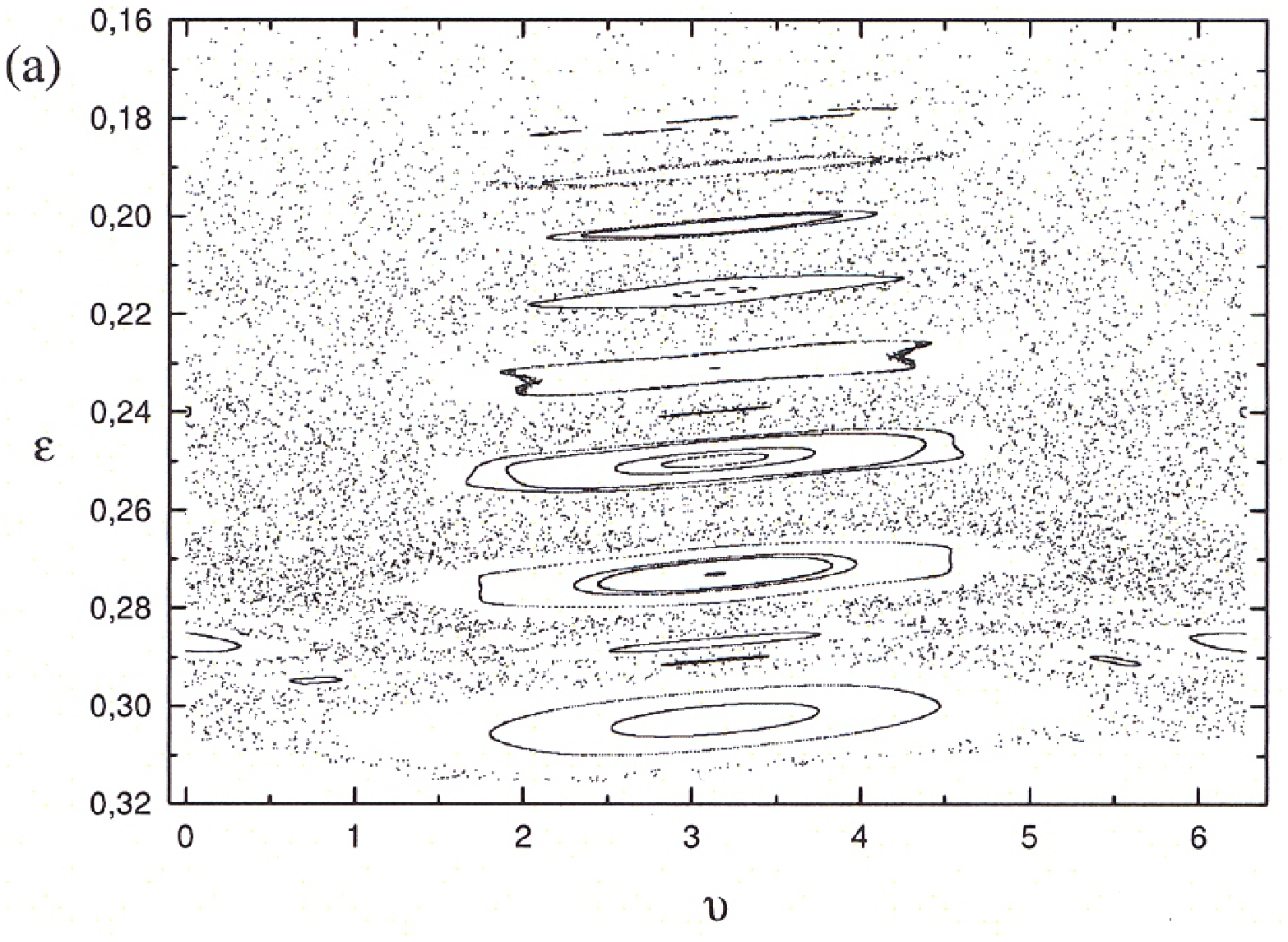} %
\includegraphics[scale=.55]{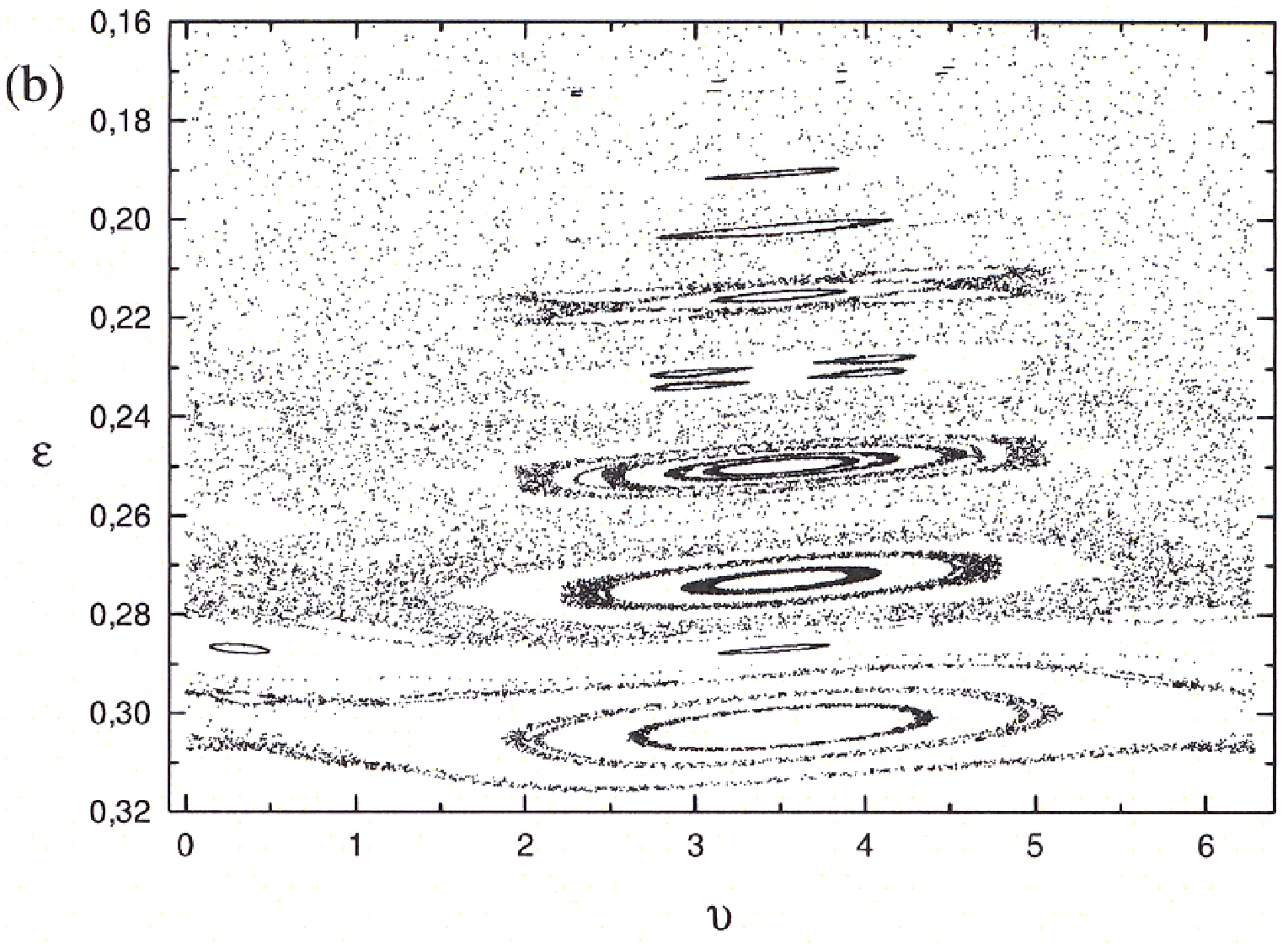}
\end{center}
\par
\vspace{-20pt}
\caption{Phase plane $(\varepsilon,\vartheta)$ of two-dimensional map
\eqref{en_skl} and \eqref{faz_skl} with parameter
$\pi F_{0}\varepsilon_{0}^{2}=0.0035$, initial conditions
$\vartheta_{0}=\pi$, $\varepsilon_{0}=0.3-0.003i$ ($i=0,1,2,\ldots$)
and orientation angle (a) $\varphi=0$ and (b) $\varphi=\pi/6$.}
\label{faz_erdv_dv}
\end{figure*}

\begin{figure*}[htb]
\begin{center}
\includegraphics[scale=.7]{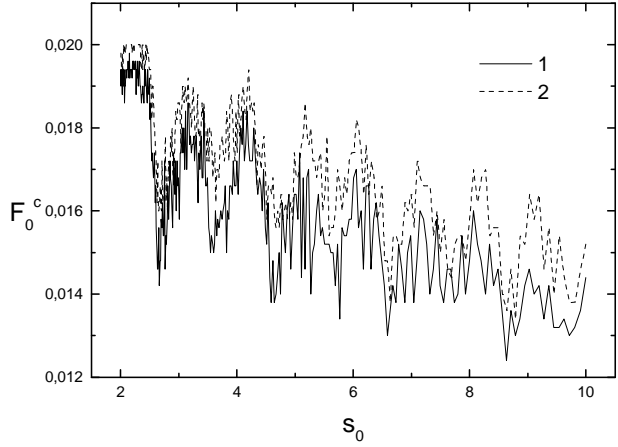}
\end{center}
\par
\vspace{-20pt}
\caption{Ionization threshold field dependence on the relative frequency
$s_{0}$ for two-dimensional atom with the eccentricity $e=0.9$ and
orientation angle $\varphi=0$ (curve~1) and $\varphi=\pi/6$ (curve~2).}
\label{kr_lauk_dv}
\end{figure*}

Using mapping equations \eqref{en_skl} and \eqref{faz_skl} we can represent
the energy and phase dynamics (figure~\ref{faz_erdv_dv}) and calculate
numerically the relative threshold field strength for the ionization of the
two-dimensional hydrogen atom (figure~\ref{kr_lauk_dv}).

\subsection{Limiting cases for energy and angular momentum changes}

\subsubsection{Approximations for very extended orbits}

For very extended orbits, $e\rightarrow 1$, the expansions for the energy %
\eqref{e-pipi} and angular momentum \eqref{m-pipi} in powers of $\beta =%
\sqrt{1-e^{2}}\ll 1$ are
\begin{multline}
\varepsilon _{j+1}=\varepsilon _{j}+\frac{4\pi F_{0}\varepsilon _{0}^{2}}{%
\varepsilon _{j+1}}\left\{ -\left( 1+\beta ^{2}\right) \mathbf{J}%
_{s}^{\prime }(s)\sin \vartheta _{j}\cos \varphi \right.  \\
\left. +\beta \left[ \mathbf{J}_{s}(s)-\frac{\sin (s\pi )}{s\pi }\right]
\cos \vartheta _{j}\sin \varphi \right\} \;,  \label{es1}
\end{multline}
\begin{multline}
\mu _{j+1}=\mu _{j}+\frac{4\pi F_{0}\varepsilon _{0}^{2}}{\varepsilon _{j+1}}%
\left\{ \beta \left[ \mathbf{J}_{s}(s)-\frac{\sin (s\pi )}{s\pi }\right]
\sin \vartheta _{j}\cos \varphi \right.  \\
\left. +\left[ -\left( 1+\beta ^{2}\right) \mathbf{J}_{s}^{\prime
}(s)+\left( 2-\beta ^{2}\right) \frac{\sin (s\pi )}{\pi }\right] \cos
\vartheta _{j}\sin \varphi \right\} .\;
\end{multline}

For $e=1$ eq. (15) coincides with eq. (4) for the one-dimensional atom,
while eq. (16) represents change of the angular momentum, resulting to
transition to the elliptic states with nonzero angular momentum.

\subsubsection{Approximations for almost circular orbits}

For almost circular orbits the eccentricity is small, $e\rightarrow 0$.
Expansion of expressions \eqref{e-pipi} and \eqref{m-pipi} in powers of $e$
may be obtained from equations (3) and (11). The results are

\begin{multline}
\varepsilon _{j+1}=\varepsilon _{j}+\frac{4F_{0}\varepsilon _{0}^{2}\sin
(s\pi )}{\varepsilon _{j+1}}\left\{ \left[ \frac{e}{2}+\frac{\frac{3}{4}%
e^{2}s^{2}-1}{1-s^{2}}+\frac{\frac{1}{2}es^{2}}{4-s^{2}}-\frac{\frac{3}{4}%
e^{2}s^{2}}{9-s^{2}}\right] \sin \vartheta _{j}\cos \varphi \right. \\[0.5cm]
\left. +s\left[ \frac{1-\frac{1}{4}e^{2}(s^{2}-4)}{1-s^{2}}-\frac{e}{4-s^{2}}%
+\frac{\frac{1}{4}e^{2}s^{2}}{9-s^{2}}\right] \cos \vartheta _{j}\sin
\varphi \right\} \;,  \label{es2}
\end{multline}
\begin{multline}
\mu _{j+1}=\mu _{j}+\frac{4F_{0}\varepsilon _{0}^{2}\sin (s\pi )}{%
\varepsilon _{j+1}} \\
\times \left\{ s\left[ -\frac{e}{2}+\frac{1-e^{2}(1+\frac{1}{4}s^{2})}{%
1-s^{2}}+\frac{e(1-\frac{1}{2}s^{2})}{4-s^{2}}+\frac{\frac{1}{4}e^{2}s^{2}}{%
9-s^{2}}\right] \sin \vartheta _{j}\cos \varphi \right. \\[0.5cm]
\left. +\left[ \frac{3e}{2}+\frac{s^{2}(\frac{1}{2}e^{2}(1+\frac{1}{2}%
s^{2})-1)}{1-s^{2}}-\frac{\frac{1}{2}es^{2}}{4-s^{2}}+\frac{\frac{1}{2}%
e^{2}s^{2}(3-\frac{1}{2}s^{2})}{9-s^{2}}\right] \cos \vartheta _{j}\sin
\varphi \right\} \;.
\end{multline}
These expansions are not valid for $s=1,2,3,...$.

\section{Atom in the circularly polarized field}

The equation for the energy change of the atom in the circularly polarized
field is
\begin{equation}
\dot{E_{k}}=-eF\left( v_{x}\cos (\omega t+\vartheta )\pm v_{y}\sin (\omega
t+\vartheta )\right) \;.  \label{en_kit_apskr}
\end{equation}
Here sign $"+"$ or $"-"$ corresponds to the same and opposite directions of
the electron and field rotations, respectively.

For the electron, moving between two subsequent pericenter the energy and
angular momentum changes are
\begin{equation}
\varepsilon _{j+1}=\varepsilon _{j}-\frac{4\pi F_{0}\varepsilon _{0}^{2}}{%
\varepsilon _{j+1}}\left\{ \mathbf{J}_{-s}^{\prime }(z)\pm \frac{\sqrt{%
1-e^{2}}}{e}\left[ \mathbf{J}_{-s}(z)-\frac{\sin (s\pi )}{s\pi }\right]
\right\} \sin (s\pi +\vartheta _{j}\mp \varphi )\;,  \label{e02pia}
\end{equation}
\begin{multline}
\mu _{j+1}=\mu _{j}+\frac{4\pi F_{0}\varepsilon _{0}^{2}}{\varepsilon _{j+1}}%
\left\{ \left[ \frac{\sqrt{1-e^{2}}}{e}\left( \mathbf{J}_{-s}(z)-\frac{\sin
(s\pi )}{s\pi }\right) \right. \right. \\
\left. \left. \mp \left( -\mathbf{J}_{-s}^{\prime }(z)+(e-1)\frac{\sin (s\pi
)}{\pi }\right) \right] \sin (s\pi +\vartheta _{j}\mp \varphi )\right. \\
\left. \pm \frac{(1+1/e)\cos (s\pi )+1-1/e}{\pi }\cos (s\pi +\vartheta
_{j}\mp \varphi )\right\} \;.  \label{m02pia}
\end{multline}

It should be noted, that expression in the curly brackets in eq. %
\eqref{e02pia} coincides with the expression in the quasiclassical radial
dipole matrix element in the velocity representation \cite{k1}
\begin{equation}
D_{p}^{\pm }=\frac{1}{s}\left\{ \mathbf{J}_{-s}^{\prime }(z)\pm \frac{\sqrt{%
1-e^{2}}}{e}\left[ \mathbf{J}_{-s}(z)-\frac{\sin (s\pi )}{s\pi }\right]
\right\} .
\end{equation}

This correspondence, however, takes place only for interaction of the
hydrogen atom with the circularly polarized microwave field and for
integration of the equations of motion between two subsequent pericenters.
In general, the energy \cite{gk,k2} and angular momentum changes depend on
the integration interval. So, for motion of the electron between two
subsequent apocenters, i.e., the most distant from the nucleus points, where
the electron's energy change is minimal, the energy change is described by
the expression similar to eq. \eqref{e02pia} but instead of $\mathbf{J}%
_{-s}(es)$ and $\mathbf{J}_{-s}^{\prime }(es)$ we have the Anger function
and its derivative of the positive order, $\mathbf{J}_{s}(es)$ and $\mathbf{J%
}_{s}^{\prime }(es)$. This interval has been used in \cite{gk,kv,kgv} for
derivation of the Kepler map for the one-dimensional hydrogen atom.

\section{Conclusion}

Analytical expressions for the energy and angular momentum changes of the
two-dimensional hydrogen atom in linearly and circularly polarized
electromagnetic fields are derived. It should be noted that in general the
expressions are rather complicated. The approximate expressions for limiting
cases of the parameters are more convenient for analytical and numerical
analysis of the dynamics.

The derived expressions are suitable for the three-dimension hydrogen atom
as well, and may be generalized for analysis of the chaotic motion (due to
the Jupiter perturbations) of comets and asteroids in the Sun system.

\end{document}